\documentclass[12pt,a4paper]{article}
\pdfoutput=1
\usepackage{cite}
\usepackage{array}
\usepackage{epsfig}
\usepackage{amssymb}
\usepackage{graphics,graphpap}
\usepackage{amsmath}
\usepackage{mcite}
\usepackage{slashed}
\usepackage{url}
\usepackage{booktabs}
\usepackage[usenames,dvipsnames]{xcolor}

\newcommand{\be}{\begin{equation}}
\newcommand{\ee}{\end{equation}}
\newcommand{\bea}{\begin{eqnarray}}
\newcommand{\eea}{\end{eqnarray}}

\newcommand{\bi}{\begin{itemize}}
\newcommand{\ei}{\end{itemize}}

\newcommand{\re}{\mathrm{Re}\,}

\newcommand{\UCha}{U}
\newcommand{\VCha}{V}
\newcommand{\ZNeu}{N}
\newcommand{\neu}[1]{\tilde{\chi}^0_{#1}}
\newcommand{\cha}{\tilde{\chi}}

\newcommand{\mneu}[1]{m_{\tilde{\chi}^0_{#1}}}
\newcommand{\mcha}[1]{m_{\tilde{\chi}^\pm_{#1}}}

\newcommand{\gev}{\ \mathrm{GeV}}
\newcommand{\fb}{\ \mathrm{fb}}

\let\oldmarginpar\marginpar
\renewcommand\marginpar[1]{\-\oldmarginpar[\raggedleft\footnotesize #1]%
{\raggedright\footnotesize #1}}

\newcommand\T{\rule{0pt}{2.5 ex}}

\newcommand\B{\rule[-1.5ex]{0pt}{0pt}}

\begin{document}
\begin{titlepage}
\renewcommand{\thefootnote}{\fnsymbol{footnote}}
\begin{flushright}
DESY 12-207
\end{flushright}

\begin{center}
{\large \bf \boldmath
One-loop effects on MSSM parameter determination\\[.4em] via chargino production at the
LC}
\vskip 0.55cm 
{\sc
Aoife Bharucha\footnote{aoife.bharucha@desy.de}$^{,1}$,
Jan Kalinowski\footnote{jan.kalinowski@fuw.edu.pl}$^{,2}$,
Gudrid Moortgat-Pick\footnote{gudrid.moortgat-pick@desy.de}$^{,1,3}$,
Krzysztof Rolbiecki\footnote{krzysztof.rolbiecki@desy.de}$^{,3,4}$
and  Georg~Weiglein\footnote{georg.weiglein@desy.de}$^{,3}$
} \vskip 0.55cm
\begin{small} 
        $^1$ {\em
II. Institut f\"{u}r Theoretische Physik, University of Hamburg, Luruper
Chaussee 149, D-22761 Hamburg, Germany}\\
\vskip 0.1cm
        $^2$ {\em Faculty of Physics, University of Warsaw, 00681 Warsaw,
Poland}\\
\vskip 0.1cm
$^3$ {\em DESY, Deutsches Elektronen-Synchrotron, Notkestr. 85, D-22607 Hamburg,
Germany}\\
\vskip 0.1cm
$^4$ {\em Instituto de F\'{\i}sica Te\'{o}rica, IFT-UAM/CSIC, 28049 Madrid,
Spain.}\\
\vskip 0.55cm
\end{small}
{\large\bf Abstract\\[9.pt]}\parbox[t]{\textwidth}{
At a future linear collider, very precise measurements, typically with errors of
$<1\%$, are expected to be achievable. Such an accuracy yields sensitivity to
quantum corrections, which therefore must be incorporated into theoretical
calculations in order to determine the underlying new physics parameters from linear
collider measurements. In the context of the chargino--neutralino sector of the
minimal supersymmetric standard model, this involves fitting one-loop
predictions to prospective measurements of cross sections, forward-backward
asymmetries and the accessible chargino and neutralino masses. Taking
recent results from LHC SUSY and Higgs searches into account, we consider three phenomenological scenarios,
each displaying characteristic features.  Our analysis demonstrates how an accurate
determination of the desired parameters is possible, and could additionally provide
access to the stop masses and mixing angle.
} \vfill

\end{center}
\end{titlepage}

\setcounter{footnote}{0}

\section{Introduction}\label{sec:1}

A linear collider (LC)~\cite{AguilarSaavedra:2001rg,Abe:2001gc,Abe:2001wn,BrauJames:2007aa,Djouadi:2007ik} will be an ideal environment for high precision studies 
of physics beyond the standard model (BSM).
A particularly well-motivated description of BSM physics is provided by the minimal supersymmetric standard 
model (MSSM), offering the lightest supersymmetric particle, often the lightest neutralino, as a candidate to explain the
evidence for dark matter in the universe~\cite{Goldberg:1983nd,Ellis:1983ew}.
Naturalness arguments (see e.g.~ref.~\cite{Hall:2011aa}) support the possibility that
higgsino-like charginos and neutralinos are light, as also predicted by
GUT motivated SUSY models~\cite{Brummer:2011yd}.
Further, due to the challenges involved in detecting electroweakinos at the LHC, 
current bounds from ATLAS and CMS only exclude small regions of 
parameter space, see e.g. refs.~\cite{Chatrchyan:2012pka,Aad:2012hba}.
Charginos and neutralinos could therefore be within reach of a first stage
linear collider.

One approach to determine the fundamental MSSM parameters is to consider constrained models
such as the constrained minimal supersymmetric standard model (CMSSM),
and perform a global fit of this reduced set of parameters to all relevant experimental results available, 
see e.g.~ref.~\cite{Fittino}.
On the other hand, in order to precisely determine the nature of the underlying 
SUSY model, we wish to determine the fundamental parameters in the most model-independent way  
possible.
The determination of the U(1) parameter $M_1$, the SU(2) parameter $M_2$, the
higgsino 
parameter $\mu$ and $\tan\beta$, the ratio of the vacuum expectation values of
the two neutral 
Higgs doublet fields, at the
percent level via chargino and neutralino pair-production has been shown to be
possible at LO (see ref.~\cite{ILCDBD} and references therein).
However, due to the expected high precision of mass and coupling measurements at the LC, 
as well as the fact that one-loop effects in the MSSM may be sizeable, higher 
order effects have to be considered.
Taking these corrections into account additional MSSM parameters become relevant,
such as the masses of the stops and sleptons, which, like the electroweakinos, are also weakly
constrained by the LHC.

In this paper we demonstrate that the determination of the fundamental 
parameters of the chargino and neutralino sector at the LC would be possible, 
even on including the complications arising due to higher order effects. 
This requires the calculation of the next-to-leading order (NLO) corrections to the
cross-section ($\sigma$) and forward-backward asymmetry ($A_{FB}$) for chargino
production, as well as to the chargino and neutralino masses.
A number of next-to-leading order (NLO) calculations of chargino and neutralino 
pair production at the LC can be found in the literature~\cite{Oller:2005xg,Fritzsche:2004nf,Kilian:2006cj,Robens:2008sa,bfmw}.
We perform our calculations in the on-shell (OS) scheme such that, as far as possible, the 
mass parameters can be interpreted as the physical masses.
Recent work on the OS renormalization of the chargino-neutralino sector can be
found 
in refs.~\cite{Fritzsche:2005,Fowler:2009ay,AlisonsThesis,Chatterjee:2011wc,Heinemeyer:2011gk,Bharucha:2012re,bfmw}. 

By fitting these loop corrected predictions to the projected experimental results, we show that
it would be possible to extract the fundamental parameters of the MSSM Lagangian.
However due to the greater number of parameters, performing the fit is more
involved than for the LO analysis.
Choosing three potential MSSM scenarios, we assess the impact of the loop
corrections and the feasibility of such an extraction in each.
We further investigate the impact of obtaining masses of the charginos and 
neutralinos from threshold scans rather than the continuum (see ref.~\cite{AguilarSaavedra:2001rg}) 
on the resulting accuracy of the parameters obtained from the fit.

The paper is organised as follows.
In sec.~\ref{sec:tree} we introduce the process studied
and define necessary notation.
We then provide details of the calculation of the loop corrections in
sec.~\ref{sec:one-loop}, including details of the renormalization scheme used.
In sec.~\ref{sec:num-res} we further discuss the method employed in order
to fit to the MSSM parameters, define the scenarios considered, and present our results.
Finally in sec.~\ref{sec:conc} we discuss the implications of the
results of the fits.
\section{Process studied and tree-level relations}\label{sec:tree}
In this paper we study the determination of the fundamental parameters in
the chargino--neutralino sector of the MSSM, via chargino production at a LC.
The charginos, $\tilde{\chi}^\pm$, and neutralinos, $\tilde{\chi}^0$, are the mass eigenstates of the gauginos and
higgsinos, as seen from the relevant part of the MSSM Lagrangian~\cite{Haber:1984rc},
\begin{align}
\nonumber\mathcal{L}_{\tilde{\chi}} = &\overline{\tilde{\chi}^-_i}(
\displaystyle{\not}p
\,\delta_{ij}-P_L (U^{*}X V^{\dagger})_{ij}-P_R (V X^\dagger
U^{T})_{ij})\tilde{\chi}^-_j\\
&+\frac{1}{2} \overline{\tilde{\chi}^0_i}( \displaystyle{\not}p\, 
\delta_{ij}-P_L (N^{*}Y N^{\dagger})_{ij}-P_R (N Y^\dagger
N^{T})_{ij})\tilde{\chi}^0_j,
\end{align}
where $P_{L/R}=1/2(1\mp\gamma_5)$. The mass matrix for the charginos is
given by 
\begin{equation}\label{eq:X}
X=
\left( \begin{array}{cc}
M_2 & \sqrt{2} M_W s_\beta  \\
\sqrt{2} M_W c_\beta  & \mu
\end{array} \right),
\end{equation}
where $s_\beta/c_\beta\equiv\sin\beta/\cos\beta$, and $M_W$ is the mass of the $W$
boson. 
This matrix is diagonalised via the
bi-unitary transformation $\displaystyle\mathbf{M}_{\tilde{\chi}^+}=U^* X
V^\dag$,
where $U$ and $V$ are complex unitary matrices.
The mass matrix for the neutralinos in the
$(\tilde{B},\tilde{W},\tilde{H}_1,\tilde{H}_2)$ basis is given by
\begin{equation}\label{eq:Y}
Y =\left( \begin{array}{cccc}
M_1 & 0 & -M_Z c_\beta s_W & M_Z s_\beta s_W \\
0   & M_2 & M_Z c_\beta c_W & -M_Z s_\beta c_W \\
-M_Z c_\beta s_W & M_Z c_\beta c_W & 0 & -\mu \\
M_Z s_\beta s_W & -M_Z s_\beta c_W & -\mu & 0 \end{array} \right),
\end{equation}
where $s_W(c_W)$ is the $\sin(\cos)$ of the weak mixing
angle $\theta_W$.
Since $Y$ is complex symmetric, its
diagonalisation requires only one unitary matrix $N$, via
$\displaystyle\mathbf{M}_{\tilde{\chi}^0}=N^*Y N^\dag$.

\begin{figure}
\begin{center}
\includegraphics[scale=0.83]{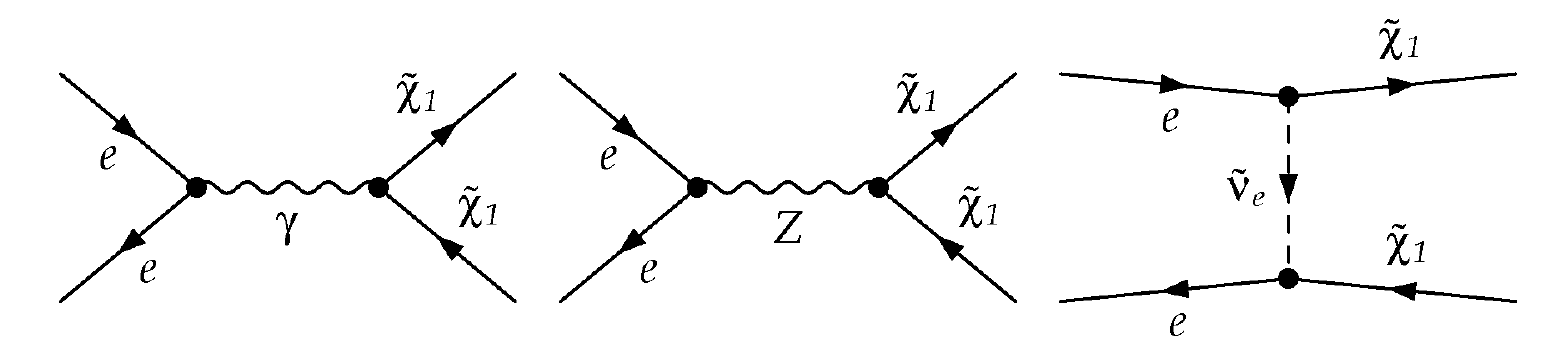}
\caption{Tree-level diagrams for the production of charginos
$\tilde\chi^+_{1}$ and $\tilde\chi^-_{1}$ at the LC.\label{fig:eeX1X1tree}}
\end{center}
\end{figure}

As described in detail in sec.~\ref{sec:num-res}, in addition to the experimental determination of 
the masses of the charginos and neutralinos the parameter determination 
relies on the measurement of the polarised cross-section for 
the pair production of charginos, $\tilde{\chi}^-_1$,
\begin{equation}
\sigma(e^+ e^- \to \tilde{\chi}^{+}_1 \tilde{\chi}^-_1),
\label{eq_proc}
\end{equation}
and the forward-backward asymmetry defined by,
\begin{equation}
 A_{FB}=\frac{\sigma(\cos\theta>0)-\sigma(\cos\theta<0)}{\sigma(\cos\theta>0)+\sigma(\cos\theta<0)},
\end{equation}
for the unpolarised cross-section, where $\theta$ is the angle of the momentum of the chargino $\tilde{\chi}^-_1$
with respect to the momentum of the incoming electron $e^-$.
Neglecting the electron-Higgs couplings, the relevant process occurs at leading order via three 
diagrams, as seen in
fig.~\ref{fig:eeX1X1tree}.

The
transition matrix element can be written as~\cite{Choi:2000ta},
\begin{equation}
 \mathcal
M_{\alpha\beta}(e^+e^-\to\tilde\chi^+_i\tilde\chi^-_j)=\frac{e}{s}Q_{\alpha\beta
}\left[\bar v(e^+)\gamma_\mu P_\alpha u(e^-)\right]\left[\bar
u(\tilde\chi^-_j)\gamma^\mu P_\beta v(\tilde\chi^+_i)\right],\label{eq:transAmp}
\end{equation}
where
$Q_{\alpha\beta}$ denotes the bilinear charges, $\alpha=L,R$ 
refers to the chirality of the $e^+e^-$ current and $\beta=L,R$ to that of the 
$\tilde{\chi}_i^+\tilde{\chi}_j^-$ current. The summation over $\alpha$
and $\beta$ is implied. In our notation, the bilinear charges are comprised of the propagators and
couplings 
\begin{align}
\nonumber Q_{LL}=&\,C^L_{\tilde\chi^+_i\tilde\chi_j^-\gamma}-D_Z G_L
C^L_{\tilde\chi^+_i\tilde\chi_j^-Z},\\
\nonumber Q_{RL}=&\,C^L_{\tilde\chi^+_i\tilde\chi_j^-\gamma}-D_Z G_R
C^L_{\tilde\chi^+_i\tilde\chi_j^-Z},\\
\nonumber Q_{LR}=&\,C^R_{\tilde\chi^+_i\tilde\chi_j^-\gamma}+D_Z G_L
\left(C^{R}_{\tilde\chi^+_i\tilde\chi_j^-Z}\right)^*+\frac{i}{2\,e}D_{\tilde\nu}
\left(C^{R}_{\tilde\nu_e e^+\tilde\chi_i^-}\right)^*C^R_{\tilde\nu_e
e^+\tilde\chi_j^-},\\
Q_{RR}=&\,C^R_{\tilde\chi^+_i\tilde\chi_j^-\gamma}+D_ZG_R
\left(C^{R}_{\tilde\chi^+_i\tilde\chi_j^-Z}\right)^*,
\end{align}
for which the required MSSM couplings for the
$\tilde\chi^+_{i}\tilde\chi^-_{j}\gamma$, $\tilde\chi^+_{i}\tilde\chi^-_{j}Z$
and
$e\tilde\nu_e\tilde\chi^+_{i}$ vertices are given by
\begin{align}
\nonumber C^{L/R}_{\tilde\chi^+_i\tilde\chi_j^-\gamma}=&\,i e\delta_{ij},\\
\nonumber C^L_{\tilde\chi^+_i\tilde\chi_j^-Z}=&\,-\frac{i e}{c_W s_W}\left(
s_W^2\delta_{ij}-U^*_{j1}U_{i1}-\frac{1}{2}U^*_{j2}U_{i2}\right),\\
\nonumber
C^R_{\tilde\chi^+_i\tilde\chi_j^-Z}=&\,C^L_{\tilde\chi^+_i\tilde\chi_j^-Z}(U\to
V^*),\\
C^R_{\tilde\nu_e e^+\tilde\chi_i^-}=&-\frac{i e}{s_W}V_{i1},
\end{align}
and $G_L$, $G_R$, $D_Z$ and $D_{\tilde\nu}$ are defined via
\begin{align}
\nonumber G_L=& \frac{s_W^2-\tfrac{1}{2}}{s_W c_W},& G_R=&\frac{s_W}{c_W}\, ,
\\
D_Z=&\frac{s}{s-M_Z^2},& D_{\tilde\nu}=&\frac{s}{t-m_{\tilde \nu}^2}. 
\end{align}
In the equations above, $e$ denotes 
the electric charge, $m_e$ and $M_Z$ are the masses of the electron and $Z$
boson.
$D_Z$ and $D_{\tilde\nu}$ refer to the propagators of the $Z$ boson 
and sneutrino (of mass $m_{\tilde \nu}$), in terms of the Mandelstam variables 
$s$ and $t$. 

Therefore in order to express the transition matrix element eq.~(\ref{eq:transAmp}) at leading order, 
the necessary MSSM parameters are $M_2$, $\mu$, $\tan\beta$ and $m_{\tilde \nu}$. However, the expected accuracy of 
the measurements at the linear collider is such that 
one-loop corrections become relevant, and we shall see in the following section 
that at higher orders eq.~(\ref{eq:transAmp}) depends on many additional MSSM parameters.

\section{NLO contributions and renormalization}
\label{sec:one-loop}
\begin{figure}[htb]
\begin{center}
\includegraphics[scale=0.8]{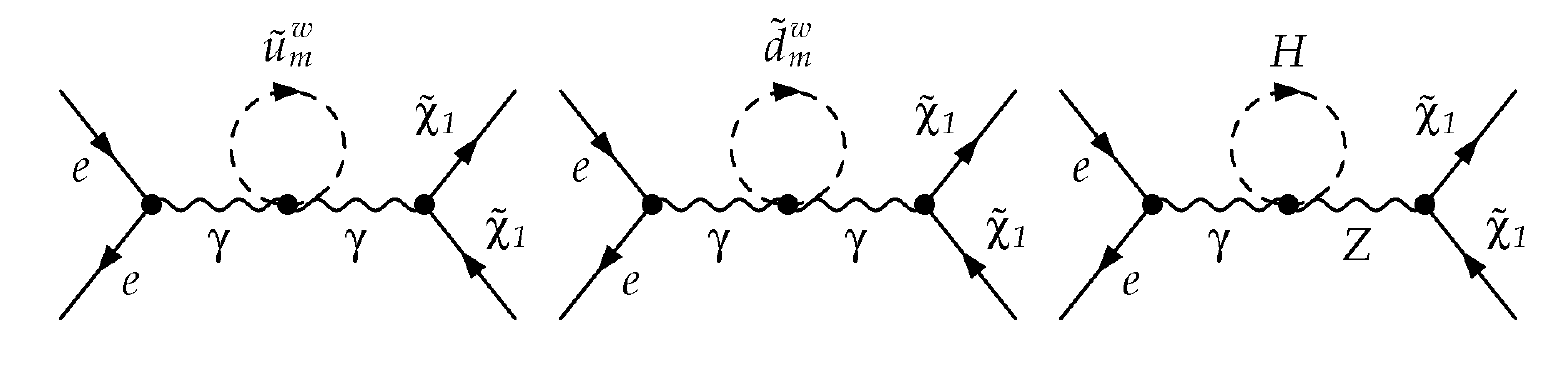}\\
\includegraphics[scale=0.8]{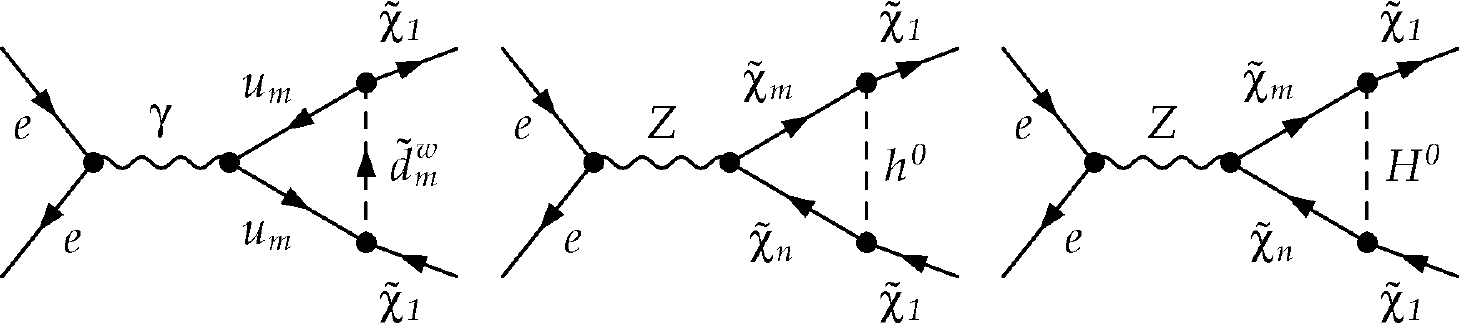}\\
\includegraphics[scale=0.8]{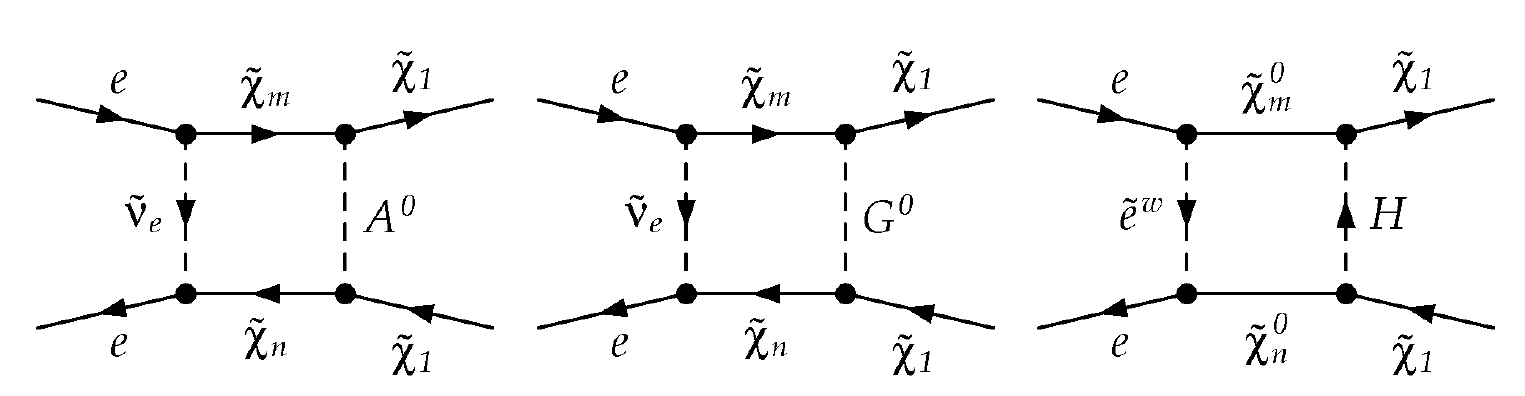}
\caption{Examples of one-loop self-energy (upper), vertex (middle) and box 
(lower) diagrams for the production of charginos $\tilde\chi^+_{1}$ and
$\tilde\chi^-_{1}$ at the LC.\label{fig:eeXXloopDiagrams}}
\end{center}
\end{figure}

We have calculated the full one-loop corrections to the forward-backward asymmetry 
for process $e^+e^-\to \tilde{\chi}^+_1\tilde{\chi}^-_1$, within
the complex MSSM; the corresponding corrections to the cross section were calculated
in ref.~\cite{bfmw}. Examples for the contributing self-energy, vertex and box diagrams are
shown in fig.~\ref{fig:eeXXloopDiagrams}. As in ref.~\cite{bfmw}, for the calculation we
have used the program
\texttt{FeynArts}~\cite{Kublbeck:1990xc,Denner:1992vza,FAorig,Hahn:2000kx,
Hahn:2001rv},
which allowed an automated generation of the Feynman diagrams and
amplitudes. Together with the packages 
\texttt{FormCalc}~\cite{Hahn:1998yk,FormCalc2,FormCalc3} and 
\texttt{LoopTools}~\cite{Hahn:1998yk} we derived
the final matrix elements and loop integrals.
We assume a unit CKM matrix.
We regularise using dimensional
reduction~\cite{DRED,DRED2,0503129}, which ensures that SUSY is 
preserved, via the implementation described in 
refs.~\cite{Hahn:1998yk,delAguila:1998nd}.

A number of one-loop calculations in the
gaugino-higgsino sector can be found in the literature,
mainly in the CP-conserving
MSSM~\cite{Lahanas:1993ib,Pierce:1993gj,Pierce:1994ew,Eberl:2001eu,
Fritzsche:2002bi,Oller:2003ge,Oller:2005xg,Drees:2006um,Schofbeck:2006gs,*Schofbeck:2007ib}, 
but some of which apply a renormalization scheme that is also applicable for complex 
parameters~\cite{Oller:2005xg,Drees:2006um}. CP-odd observables have 
also been calculated at the one-loop level, for
instance in refs.~\cite{Rolbiecki:2007se,Eberl:2005ay,Osland:2007xw}, but no
dedicated renormalization scheme was required in these cases as the
observables studied were UV-finite. Since we intend to extend 
the current study to the case of complex parameters, we follow 
the approach of refs.~\cite{Fowler:2009ay,bfmw} closely, where a dedicated
on-shell renormalization scheme for the chargino and neutralino sector of the MSSM with complex 
parameters was developed. In the following we will therefore only briefly discuss the parameter renormalization of 
the chargino and neutralino sector, relevant for the definitions of the parameters at loop level, and for
further details about the chargino field renormalization and the renormalization of other sectors 
refer the reader to refs.~\cite{Fowler:2009ay,AlisonsThesis,bfmw,Bharucha:2012re}.

The mass matrix in the chargino sector, eq.~(\ref{eq:X}), is renormalized via
\begin{equation}
X \to X+\delta X,
\label{eq_char_count}
\end{equation}
where $\delta X$ is defined by
\begin{equation}\label{eq:delX}
\delta X=
\left( \begin{array}{cc}
\delta M_2 & \sqrt{2} \delta (M_W s_\beta)  \\
\sqrt{2} \delta (M_W c_\beta)  & \delta \mu
\end{array} \right),
\end{equation}
and contains the renormalization 
constants $\delta M_2$ and 
$\delta \mu$, as well as renormalization constants (RCs) from other sectors, 
$\delta c_\beta$, $\delta s_\beta$ (which can be expressed in terms of $\delta \tan\beta$),
and $\delta M_W$, defined in ref.~\cite{bfmw}.
The neutralino mass matrix, eq.~(\ref{eq:Y}), is similarly renormalized via
\begin{equation}
Y \to Y+\delta Y,
\label{eq_neut_count}
\end{equation}
where $\delta Y$ is defined  analogously to $\delta X$ in eq.~(\ref{eq:delX})
and
contains the additional RC $\delta M_1$, cf.~eq.~(\ref{eq:Y}).

Following e.g.~ref.~\cite{Fowler:2009ay}, $\delta M_1$, $\delta M_2$
and $\delta \mu$ are determined by choosing three out of the total six
physical masses of the charginos and neutralinos to be on-shell,
i.e.~the tree-level masses, $m_{\tilde{\chi}_i}$, coincide with the
one-loop renormalized masses, $M_{\tilde{\chi}_i}=m_{\tilde{\chi}_i}+\Delta m_{\tilde{\chi}_i}$,
\begin{align}
\nonumber \Delta
m_{\tilde{\chi}_i}&\equiv-\frac{m_{\tilde{\chi_i}}}{2}\mathrm{Re}[\hat{\Sigma}
^L_{ii}(m_{\tilde{\chi}_i}^2)+\hat{\Sigma}^{R}_{ii}(m_{\tilde{\chi}_i}^2)]-\frac
{1}{2}\mathrm{Re}[\hat{\Sigma}^{SL}_{ii}(m_{\tilde{\chi}_i}^2)+\hat{\Sigma}^{SR}
_{ii}(m_{\tilde{\chi}_i}^2)]&\\
\label{eqn:deltami}&=0.&
\end{align}
We define the coefficients 
$\Sigma^{L/R}_{ij}(p^2)$ and $\Sigma^{SL/SR}_{ij}(p^2)$ of the self energy via
\begin{equation}
 \Sigma_{ij}(p^2)=\displaystyle{\not}p\, P_L 
\Sigma^L_{ij}(p^2)+\displaystyle{\not}p\, P_R  
\Sigma^R_{ij}(p^2)
+P_L  \Sigma^{SL}_{ij}(p^2)+ P_R \Sigma^{SR}_{ij}(p^2),
\label{eqn:Lorentzse}
\end{equation}
and define the left and right handed vector and scalar coefficients of the
renormalized self-energy analogously via
$\displaystyle\hat{\Sigma}^{L/R}_{ij}(p^2)$ and
$\displaystyle\hat{\Sigma}^{SL/SR}_{ij}(p^2)$ respectively.

As stated earlier, we consider the parameter renormalization as for the complex MSSM, such
that
our setup is easily adaptable for future extensions.
In ref.~\cite{AlisonsThesis,bfmw}, it was shown that in the CP violating case,
the 1-loop corrections to the phases of $M_1$ and $\mu$, i.e.~$\phi_{M_1}$ and
$\phi_\mu$ 
respectively\footnote{ We adopt the convention that $M_2$ is real.}
are UV finite, and it is argued that these phases can therefore be left
unrenormalized. Adopting this approach, one can determine expressions for 
$\delta|M_1|$, $\delta|M_2|$ and $\delta|\mu|$, depending on which three
physical masses are chosen to be on-shell.
As we have two external charginos, and in order to easily extend our setup to
the case
of $\tilde{\chi}_1^+\tilde{\chi}_2^-$ production, we assume the NCC scheme with
$\tilde{\chi}_1^{0}$, $\tilde{\chi}_1^{\pm}$ and
 $\tilde{\chi}_2^{\pm}$
on-shell~\cite{Fowler:2009ay,AlisonsThesis,Chatterjee:2011wc,bfmw}. Note that in choosing a scheme, it is desirable 
that the on-shell particles should contain significant bino, 
wino and higgsino components, in order that the $M_1$, $M_2$ and $\mu$ parameters are 
accessible~\cite{Fowler:2009ay,AlisonsThesis,Chatterjee:2011wc,bfmw}. For the above choice,
these conditions 
are satisfied for all the scenarios defined in sec.~\ref{sec:num-res}, in which
the 
lightest neutralino always has a sizeable bino-like component.
The parameters in question of the chargino mass matrix can
then be renormalized via expressions given in refs.~\cite{AlisonsThesis,Bharucha:2012re,bfmw}, which we list here for completeness,
\begin{align}
\nonumber\hspace{-.48cm}\delta |M_1|=&-\frac{1}{\re(e^{-i\phi_{M_1}}\,N_{i1}^2)F}\\
\nonumber&\,\,\,\Big((2\re(e^{-i\phi_{\mu}}N_{i3} N_{i4}) \re(U_{j1} V_{j1})
+\re N_{i2}^2 \re(e^{-i\phi_{\mu}}U_{j2} V_{j2}))C_{k}\\
\nonumber&+(\re(U_{j1} V_{j1})\re(e^{-i\phi_{\mu}}U_{k2} V_{k2})-\re(e^{-i\phi_{\mu}}U_{j2} V_{j2})\re(U_{k1} V_{k1}))N_{i}\\
\label{eq:NCCa} &-(\re N_{i2}^2\re(e^{-i\phi_{\mu}}U_{k2} V_{k2})+2\re(e^{-i\phi_{\mu}}N_{i3} N_{i4})\,\re(U_{k1} V_{k1}))C_{j}\Big),
\\
\label{eq:NCCb} \delta |M_2|=\,&\frac{1}{F}\Big(\re(e^{-i\phi_{\mu}}U_{j2} V_{j2})C_{k}-\re(e^{-i\phi_{\mu}}U_{k2} V_{k2}) C_{j}\Big),
\\
 \delta |\mu|=\,&-\frac{1}{F}\Big(\re(U_{j1} V_{j1})C_{k}-\re(U_{k1} V_{k1})C_{j}\Big), \label{eq:NCCc}
\end{align}
where $F$, $C_{i}$ and $N_{i}$ are defined by
\begin{align}
 F=\,&2\big(\re(U_{k1} V_{k1}) \re(e^{-i\phi_{\mu}}U_{j2} V_{j2})-\re(U_{j1} V_{j1})\,\re(e^{-i\phi_{\mu}}U_{k2} V_{k2})\big),\\
C_{i}=\,&\re \big[m_{\tilde{\chi}^+_i}[\Sigma^L_{\pm,ii}(m_{\tilde{\chi}^+_i}^2)+ \Sigma^R_{\pm,ii}(m_{\tilde{\chi}^+_i}^2)]+
      \Sigma^{SL}_{\pm,ii}(m_{\tilde{\chi}^+_i}^2)+\Sigma^{SR}_{\pm,ii}(m_{\tilde{\chi}^+_i}^2)\big]\nonumber\\&- 
     \sum_{\tiny\substack{j=1,2\\k=1,2}}2\delta X_{jk}
\re(\UCha_{ij}\VCha_{ik}), \\
%
N_{i}=\,&\re \big[m_{\tilde{\chi}^0_i}[\Sigma^L_{0,ii}(m_{\tilde{\chi}^0_i}^2)+\Sigma^R_{0,ii}(m_{\tilde{\chi}^0_i}^2)] +\Sigma^{SL}_{0,ii}(m_{\tilde{\chi}^0_i}^2)+ \Sigma^{SR}_{0,ii}(m_{\tilde{\chi}^0_i}^2)\big]\nonumber
\\& - \sum_{\tiny\substack{j=1,2\\k=3,4}}
       4\delta Y_{jk}\re(\ZNeu_{ij}\ZNeu_{ik}),
\label{eqn:CiNishorthand}
\end{align}
and the subscripts $\pm$ and $0$ identify the coefficients of the chargino and neutralino self-energy respectively.

\begin{figure}[tb]
\begin{center}
\includegraphics[scale=0.8]{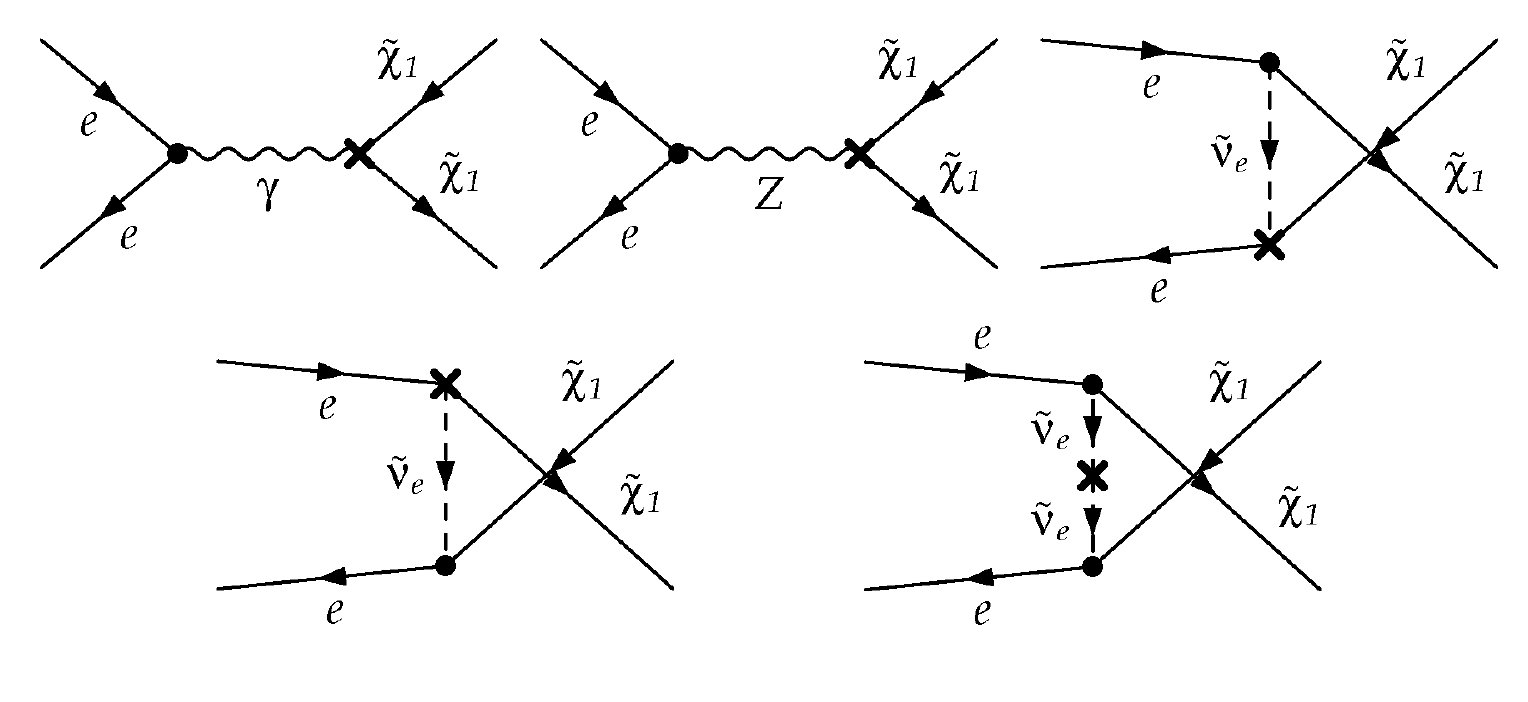}
\caption{Counterterm diagrams in the MSSM for the production of charginos
$\tilde\chi^+_{1}$
and $\tilde\chi^-_{1}$ at the LC.\label{fig:eeXXcounterDiagrams}}
\end{center}
\end{figure}

Finite results for the process of interest at one-loop are obtained by adding the counterterm
diagrams shown in fig.~\ref{fig:eeXXcounterDiagrams}. Although
\texttt{FeynArts} generates these diagrams, expressions for the
counterterms which renormalize the couplings defined at tree-level in
eq.~(\ref{eq:transAmp}), calculated in ref.~\cite{bfmw}, are required as input, and therefore, again for completeness, we
provide expressions for these explicitly.
For the $\gamma\tilde\chi^+_{i}\tilde\chi^-_{j}$,
$Z\tilde\chi^+_{i}\tilde\chi^-_{j}$ and $e\tilde\nu_e\tilde\chi^+_{i}$ vertices,
these can be expressed as follows,
\begin{equation}
\nonumber\hspace{-.4cm}\delta
C^L_{\tilde\chi^+_i\tilde\chi_j^-\gamma}=C^L_{
\tilde\chi^+_i\tilde\chi_j^-\gamma} \left(\delta Z_e+\frac{\delta
Z_{\gamma\gamma}}{2}\right)+C^L_{\tilde\chi^+_i\tilde\chi_j^-Z}\frac{\delta
Z_{Z\gamma}}{2}+\frac{i e}{2}\left(\delta Z^L_{\pm,ij}+\delta \bar Z^L_{\pm,ij}\right),
\end{equation}
\begin{align}
\nonumber\delta C^L_{\tilde\chi^+_i\tilde\chi_j^-Z}
=&\,\,C^L_{\tilde\chi^+_i\tilde\chi_j^-Z}\left(\delta Z_e-\frac{\delta
c_W}{c_W}-\frac{\delta s_W}{s_W}+\frac{\delta 
Z_{ZZ}}{2}\right)+C^L_{\tilde\chi^+_i\tilde\chi_j^-\gamma}\frac{\delta Z_{\gamma
Z}}{2}\\
&-2 i e\frac{\delta
s_W}{c_W}\delta_{ij}+\frac{1}{2}\sum_{n=1,2}\left(\delta
C^L_{\tilde\chi^+_i\tilde\chi_n^-Z}Z^L_{\pm,nj}+C^L_{
\tilde\chi^+_n\tilde\chi_j^-Z}\delta
\bar Z^L_{\pm,in}\right),
\end{align}
where the analogous right-handed parts are obtained by the replacement
$L\to R$,
and
\begin{align}
\nonumber \delta C^R_{\tilde\nu_e e^+\tilde\chi_i^-}=&\,\,C^R_{\tilde\nu_e
e^+\tilde\chi_i^-}\bigg(\delta Z_e-\frac{\delta s_W}{s_W}
+\frac{1}{2}\left(\delta Z_{\tilde \nu_e}+\delta Z^{L^*}_{e}\right)\bigg)\\
&+\frac{1}{2}\left(C^R_{\tilde\nu_e e^+\tilde\chi_1^-}\delta
Z^R_{\pm,1i}+C^R_{\tilde\nu_e e^+\tilde\chi_2^-}\delta Z^R_{\pm,2i} \right).
\end{align}
Note that the renormalization constants of the SM fields, i.e.~$Z_{VV}$ ($V=\gamma,Z$) 
and $\delta Z^{L}_{e}$ for the vector bosons and electron, and parameters, 
i.e.~$\delta Z_e$ and $\delta c_W(s_W)$ for the electric charge and $\cos(\sin)$ of the weak mixing angle respectively, 
can be found in ref.~\cite{bfmw}. The renormalization for the chargino fields is performed in the
most general manner, making use of separate RCs
for the incoming and outgoing fields, i.e.~coefficients $\delta
Z^{L/R}_{\pm,ij}$ and $\delta\bar{Z}^{L/R}_{\pm,ij}$ respectively for left
and right-handed charginos as given in ref.~\cite{bfmw}.
Finally, the counterterm for the sneutrino
self energy takes the form
\begin{equation}
\delta C_{\overline{\tilde\nu}_i \tilde\nu_j}=i
\delta_{ij}\left(\frac{1}{2}(\delta Z_{\tilde\nu_i}+\delta
Z^*_{\tilde\nu_i})p^2-\delta
m_{\tilde\nu_i}^2-\frac{m_{\tilde\nu_i}^2}{2}(\delta Z_{\tilde\nu_i}+\delta
Z^*_{\tilde\nu_i})\right), 
\end{equation}
for $\tilde\nu_i=\tilde\nu_e,\,\tilde\nu_\mu,\,\tilde\nu_\tau$, where the 
sneutrino field and mass RCs, $\delta Z^*_{\tilde\nu_i}$ and $\delta
m_{\tilde\nu_i}$, are also defined following ref.~\cite{bfmw}.

Inital and final state soft radiation must also be included to obtain an infra-red
finite result as the incoming and outgoing particles are charged, and this is
done as described in detail in ref.~\cite{bfmw}, using the phase-space slicing method 
to define the singular soft and
collinear contributions in the regions $E<\Delta E$ and $\theta<\Delta \theta$
respectively.
In the soft and collinear limit, the results are regularised using electron and photon masses, 
respectively, and factorised into analytically integrable expressions
proportional to the tree-level cross-section $\sigma^{\rm
tree}(\mathit{e}^+\mathit{e}^-\to \tilde{\chi}_1^+\tilde{\chi}_1^-)$. 
However the result is cut-off dependent (i.e.~on  $\Delta E$ and $\Delta
\theta$), and removing this dependence requires a calculation of the cross
section for the three body final state, excluding the soft and collinear
regions, which we perform using \texttt{FeynArts} and \texttt{FormCalc}.
We further require that soft photon radiation is included in the
cross-section obtained from \texttt{FormCalc}. Finally we obtain a complete IR safe 
and cut-off independent result by adding the collinear
contribution, which is calculated following the procedure outlined in
ref.~\cite{Oller:2005xg}.
\section{Fit strategy and numerical results}\label{sec:num-res}
\subsection{Obtaining MSSM parameters from the fit}

With the loop corrections calculated as in section~\ref{sec:one-loop}, we can
determine the
fundamental parameters of the MSSM at NLO. From now on, we will restrict our
study to the case of real parameters.
In the chargino and neutralino sectors there are four real parameters, see
sec.~\ref{sec:tree}, which we fit to,
\begin{eqnarray} \label{eq:parameters}
M_1, \quad  M_2, \quad \mu,  \quad \tan\beta\,.
\end{eqnarray} 
We additionally fit to the sneutrino mass, as this enters at tree level
and will therefore significantly affect cross sections and forward-backward
asymmetries. However in those scenarios where the sneutrino would already have been 
observed at the LC, its mass is assumed to be known.
At the loop level, a large number of MSSM parameters will contribute.
Depending on the scenario, only limited
knowledge about some of these may be available. In particular LHC data may only 
provide limited information about the parameters of the stop sector, and 
direct production at the LC might not be possible. 
However, our analysis also offers good sensitivity to these parameters at the
LC, 
as stops could significantly contribute to chargino/neutralino observables at
NLO.

At the LC, the accessible masses are expected to be measured with high precision using
different
methods~\cite{AguilarSaavedra:2001rg}. In the following we adopt the experimental precision which
could be achieved using the threshold scan method, however we also investigate how the
fit precision would change if the masses were obtained from the continuum.  In case of the cross
sections, the experimental uncertainty is dominated by the statistical
uncertainty~\cite{Desch:2006xp},
\begin{equation}
\frac{\Delta \sigma}{\sigma} = \frac{\sqrt{S+B}}{S},
\end{equation}
where $S$ and $B$ are the signal and background contributions,
respectively. In addition, we assume that the statistical
uncertainties for the cross sections correspond to an integrated
luminosity of ${\mathcal{L}} = 200 \fb^{-1}$ per polarisation
assuming the efficiency of $\epsilon = 15\%$, which includes branching ratios
for semileptonic final states and a selection efficiency of
$50\%$~\cite{Desch:2006xp}. 
Similarly, for the forward-backward asymmetry we have

\begin{eqnarray}
   \delta A^{\mathrm{stat}}_{FB} &=&
\sqrt{\frac{1-A_{FB}^2}{N}},
\end{eqnarray}
and the total number of events $N = N_+ + N_-$~\cite{Desch:2006xp}.

In order to estimate the theoretical uncertainty on the masses, 
cross-sections and forward backward asymmetries, we consider
the size of possible effects due either to neglected higher order corrections or to 
unknown MSSM parameters not included in the fit.
NNLO corrections are an important source of theoretical
uncertainty, however, at present, corrections of this kind are only known for 
chargino and neutralino masses, for which the leading SUSY-QCD NNLO 
corrections were calculated in ref.~\cite{Schofbeck:2006gs,*Schofbeck:2007ib}. 
Based on these results we estimate the uncertainty on the masses due to NNLO corrections to be
of the order of $0.5\gev$, i.e.~comparable to the expected experimental uncertainty. 
Note that the masses chosen on-shell are assigned no theoretical uncertainty.
We further neglect the currently unknown uncertainties arising due to NNLO 
corrections to the cross-sections and 
forward backward asymmetries, assuming that in the future NNLO results for 
these could be incorporated.
However, we do include the additional uncertainty arising due to any unknown MSSM 
parameters which are not included in the fit, dominated by the
contribution from the heavy pseudoscalar Higgs boson mass
$m_{A^0}$. We perform a multi-dimensional $\chi^2$ fit using
\texttt{Minuit}~\cite{James:1975dr,minuit}
\begin{equation}
\chi^2 = \sum_i \left| \frac{{\mathcal{O}}_i - \bar{{\mathcal{O}}}_i }
                { \delta {\mathcal{O}}_i } \right|^2 ,
\end{equation}
where the sum runs over the input observables ${\mathcal{O}}_i$, depending on
the scenario, 
with their corresponding experimental uncertainties
$\delta {\mathcal{O}}_i$. 

\subsection{Scenarios studied and motivation}
\begin{figure}[tb!]
\begin{center}
\includegraphics[scale=.85]{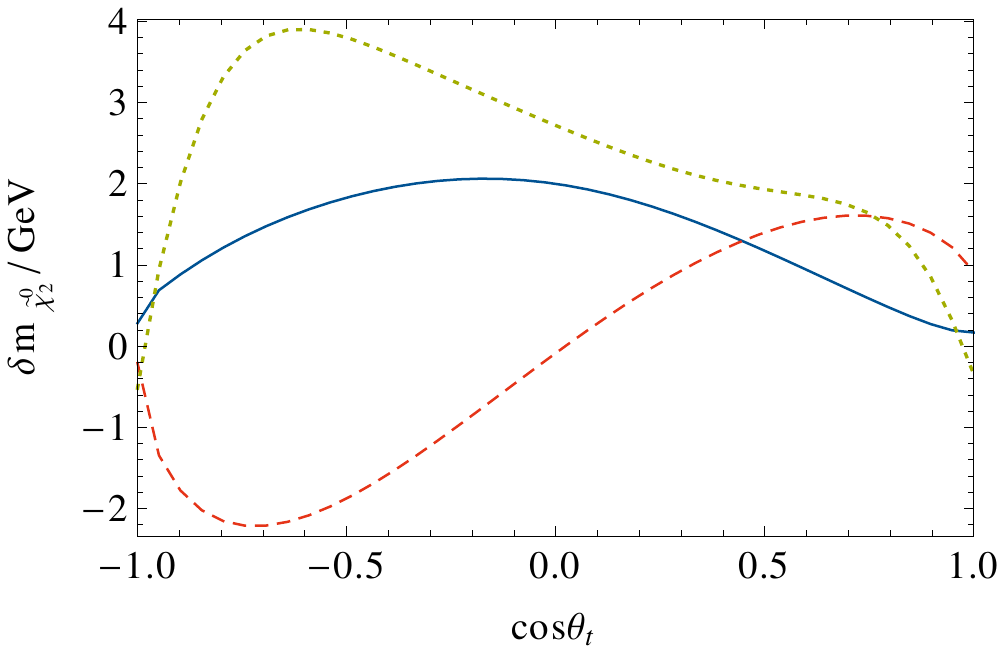}\\
\vspace{.1cm}
\includegraphics[scale=0.85]{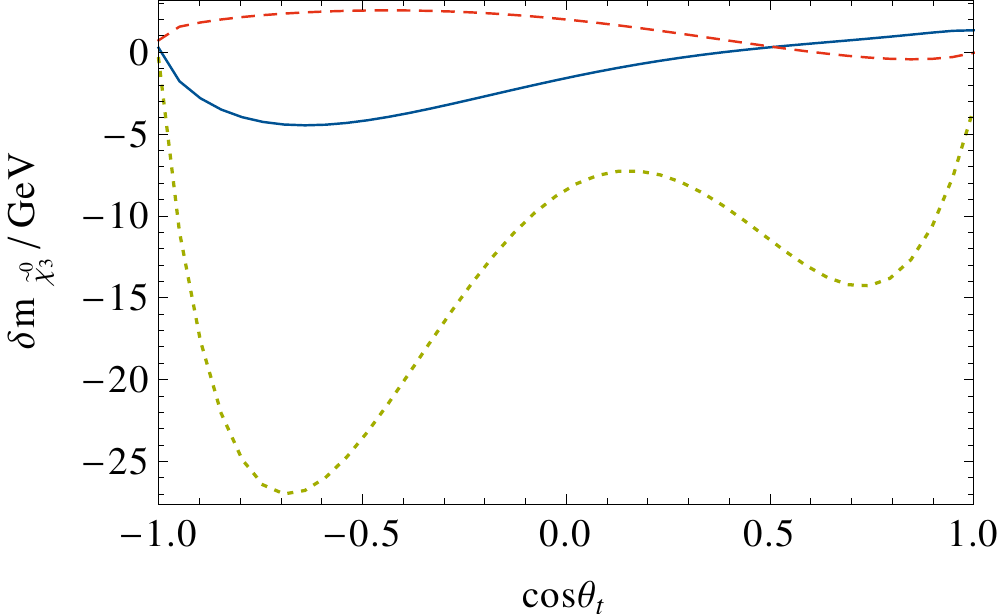}
\caption{One-loop corrections to the masses of neutralinos $\neu{2}$ (upper) and $\neu{3}$ (lower)
as a function of the stop mixing angle $\cos\theta_t$, for scenarios S1 (blue), S2 (red, dashed) and S3 (green, dotted).\label{fig:dmn-cost}}
\end{center}
\end{figure}

We carry out the fit for three scenarios, S1, S2 and S3, shown 
in tab.~\ref{tab:s1/2/3}, chosen in order to 
realistically assess the sensitivity to the desired parameters in a variety 
of possible situations.
Due to the current status of direct LHC searches~\cite{atlas:2012rz,cms:2012jx},
in all scenarios we require heavy first and second generation squarks and
gluinos, while the stop sector is assumed to be relatively light.\footnote{Note that in light of current LHC limits, 
the value $M_3$=700~GeV in S1 and S2 means that the gluino mass is rather low, however our results are largely independent of this choice
as $M_3$ only enters our calculations via two loop corrections to $m_h$.}
In S1 and S2 we take the masses
of the stops, $m_{\tilde{t}_1}$ and $m_{\tilde{t}_2}$, to be 400~GeV and 800~GeV respectively, 
and the mixing angle to be $\cos\theta_{t}=0$. The sbottom sector is then obtained by defining 
$m_{\tilde{b}_1}=400$~GeV and $\cos\theta_{b}=0$.
On the other hand in S3, in order to obtain $m_h=$ 125 GeV, calculated using \texttt{FeynHiggs
2.9.1}~\cite{Heinemeyer:1998yj,Heinemeyer:1998np,Degrassi:2002fi,Frank:2006yh}, such that it is compatible
with the recent Higgs results from the LHC~\cite{ATLAS:2012gk,CMS:2012gu},
the stop sector parameters are chosen to be $m_{u_3} = 450$~GeV, $m_{q_3} = 1500$~GeV and $A_t =   -1850$~GeV, ensuring
large mixing between the stops, such that $\cos\theta_{t}=0.148$. 
The sbottom sector is then obtained by defining 
$m_{\tilde{b}_1}=450$~GeV and $\cos\theta_{b}=0$.
In fig.~\ref{fig:dmn-cost} the mass corrections for neutralinos 
$\neu{2}$ and $\neu{3}$ are seen to be sensitive 
to the stop mixing angle for each of the studied scenarios.

As a result of indirect limits (checked using 
{\texttt{micrOmegas 2.4.1}}~\cite{Belanger:2006is, Belanger:2010gh}), we have chosen mixed gaugino higgsino
scenarios favoured by the relic density measurements~\cite{Komatsu:2010fb}
and relatively high pseudoscalar Higgs masses in light of flavour physics 
constraints, e.g.~due to the dependence of the branching ratio $\mathcal{B}(b\to s\gamma)$ on the charged Higgs mass. We also 
check that our scenarios agree with the experimental results for branching ratio 
$\mathcal{B}(b\to s\gamma)$ 
and the anomalous magnetic moment of the muon 
$\Delta(g_\mu-2)/2$.
Further, in S2 we study the sensitivity of the fit to large values of
$M_2$, 
such that the wino-like chargino and neutralino are heavy and decoupled 
from the bino and higgsino-like particles.
Finally, in S1/S2 we consider the case that the sleptons (with the exception of the light stau) and pseudoscalar Higgs 
bosons are at the TeV scale, and in S3 the case that they are relatively light.
Therefore, while S1/S2 are not in keeping with the 
125 GeV Higgs boson, they provide illustrative examples of the potential of the LC 
in scenarios complementary to S3.\footnote{Note that in S1(S2) a Higgs mass of 
$m_h=125$ GeV can also be achieved by adopting $\cos\theta_t=-0.4\,(-0.5)$.}

\begin{table}[tb!]
\begin{center}
 \begin{tabular}{lclc}
\toprule
\multicolumn{4}{c}{\T Scenario 1/2}\\
\midrule
\T$M_1$ & 125& $M_2$ & 250/2000\\
$\mu$ & 180 & $m_{A^0}$ & 1000\\
$M_3$ & 700 & $\tan\beta$ & 10\\
$M_{q_{1,2}}$ & 1500 &$A_{q_{1,2}}$ & 650\\
$M_{l/e_{1,2}}$ & 1500 &$A_{l_{i}}$ & 650\\
\B$M_{{l}_{3}}$ & 800 &$M_{e_3}$ & 400\\
\midrule
\multicolumn{4}{c}{\T Scenario 3}\\
\midrule
\T$M_1$ & 106& $M_2$ & 212\\
$\mu$ & 180 & $m_{A^0}$ & 500\\
$M_3$ & 1500 & $\tan\beta$ & 12\\
$M_{q_{1,2}}$ & 1500 & $A_{q_{1,2}}$ & -1850\\
$M_{l_{i}}$ & 180 & $A_{l_{i}}$ & -1850\\
\B $M_{e_{1,2}}$ & 125 & $M_{e_{3}}$ & 106\\
\bottomrule
\end{tabular}
\caption{Parameters for scenarios 1/2 and 3 (S1/S2 and S3), in GeV with the exception of $\tan\beta$. Here $M_{(l/q)_{i}}$
($M_{(e/u/d)_{i}}$) represent the left (right) handed mass parameters for a
slepton/squark of generation $i$ respectively, and $A_f$ is the trilinear
coupling for a sfermion $\tilde f$. See text for stop and sbottom parameter definitions. \label{tab:s1/2/3}}
\end{center}
\end{table}

\subsection{Results for scenario 1 \label{sec:scenario1}}

In this scenario, only the charginos and three neutralinos will be accessible at the
LC. As input for the fit we
therefore use:
\begin{itemize}
\item the masses of the charginos ($\cha_1^{\pm},\,\cha_2^{\pm}$) and three lightest
neutralinos ($\neu{1},\,\neu{2},\,\neu{3}$)
\item the light chargino production cross section $\sigma(\cha^+_1\cha^-_1)$
with polarised beams at $\sqrt{s} = 350$ and $500 \gev$
\item the forward-backward asymmetry $A_{FB}$ at $\sqrt{s} = 350$ and $500 \gev$
\item the branching ratio $\mathcal{B}(b\to s\gamma)$ calculated using
{\texttt{micrOmegas 2.4.1}}~\cite{Belanger:2006is, Belanger:2010gh}.
\end{itemize}
The input variables, together with errors, namely the assumed experimental precision 
of the prospective LC measurements as well as the theoretical uncertainties, are listed in tab.~\ref{tab:inputsc1}.
It is interesting to observe the large NLO corrections to $A_{FB}$, which even result in a change of sign.
Note that $\mathcal{B}(b\to s\gamma)$ is included in order to increase
sensitivity to 
the third generation squark sector, and an estimated experimental precision of $0.3\cdot 10^{-4}$, taken from
ref.~\cite{O'Leary:2010af}, is adopted. 
We found that the impact of the muon anomalous magnetic moment is negligible in this
scenario, mainly due to the heavy smuon sector.
It should be possible to probe the supersymmetric QCD sector, 
with sqark masses of $\sim$1.5~TeV and the gluino mass of $\sim 700$ GeV, at the LHC, such that the theoretical uncertainty 
arising due to these parameters is small in comparison to that due to the unknown $m_{A^0}$.
We therefore include the small dependence on the $A^0$ mass as an additional source of error,
having explicitly checked that the impact of all other parameters is negligible. Note that there are no theoretical
errors for masses chosen to be on-shell. Even at one loop, these masses are related to the fundamental
parameters via the tree level relations, and are included in the fit.
\begin{table}[tb!]
\renewcommand{\arraystretch}{1.5}
\begin{center}
\begin{tabular}{@{}lr@{.}lr@{.}lr@{.}lr@{.}l@{}}
\toprule
Observable & \multicolumn{2}{c}{Tree value} & \multicolumn{2}{c}{Loop corr.} & \multicolumn{2}{c}{Error exp.} & \multicolumn{2}{c}{Error th.}\\ 
\midrule
$\mcha{1}$ & \quad$149$ & $6$ & \multicolumn{2}{c}{$-$} & \quad$0$ & $1\ (0.2)$ & \multicolumn{2}{c}{$-$} \\
$\mcha{2}$ & $292$ & $3$ & \multicolumn{2}{c}{$-$} & 0 & $5\ (2.0)$ & \multicolumn{2}{c}{$-$} \\
$\mneu{1}$ & $106$ & $9$ & \multicolumn{2}{c}{$-$} &0 & 2 & \multicolumn{2}{c}{$-$} \\
$\mneu{2}$ & $164$ & $0$ & $2$ & $0$ & 0 & $5\ (1.0)$ & \quad\,\,$0$ & $5$ \\
$\mneu{3}$ & $188$ & $6$ & $-1$ & $5$ & 0 & $5\ (1.0)$ &0 & 5 \\

$\sigma(\cha^+_1\cha^-_1)^{350}_{(-0.8,0.6)}$ & $2347$ & $5$ & $-291$ & $3$ & 8 & 7 & 2 & 0
 \\
$\sigma(\cha^+_1\cha^-_1)^{350}_{(0.8,-0.6)}$ & $224$ & $4 $ & $7$ & $6 $ & 2 & 7 & 0 & 5
\\
$\sigma(\cha^+_1\cha^-_1)^{500}_{(-0.8,0.6)}$ & $1450$ & $6 $ & $-24$ & $4$ & 8 & 7  & 2 & 0
 \\
$\sigma(\cha^+_1\cha^-_1)^{500}_{(0.8,-0.6)}$ & $154$ & $8  $ & $ 12$ & $7 $ & 2 & 0 & 0 & 5
  \\
$A_{FB}^{350}(\%)$ & $-2$&2& 6&8&0&8&0&1\\
$A_{FB}^{500}(\%)$ &$-2$&6 &5&3&1&0&0&1\\
\bottomrule
\end{tabular}
\end{center}
\label{tab:inputsc1}
\caption{Observables (masses in GeV, cross sections in fb) used as input for the fit in S1, tree-level values and loop
corrections are specified. 
Here the superscript on $\sigma$ and $A_{FB}$ denotes $\sqrt{s}$ in GeV, and the subscript on $\sigma$ 
denotes the beam polarisation $(\mathcal{P}(e^-),\mathcal{P}(e^+))$. 
Errors in brackets are for masses obtained from the continuum.
The central value of the theoretical prediction, $\mathcal{B}(b\to s\gamma)= 3.3\cdot 10^{-4}$ GeV is also included in the fit. 
 See text for details of the calculation and error estimation.}
\end{table}

In S1 we fit 8 MSSM parameters: $M_1$, $M_2$, $\mu$, $\tan\beta$,
$m_{\tilde{\nu}}$, $\cos\theta_{t}$, $m_{\tilde{t}_1}$, and
$m_{\tilde{t}_2}$. The results of the fit
are given in tab.~\ref{tab:ressc1}.
We find that the gaugino and higgsino mass parameters are determined with an 
accuracy better than 1\%, while $\tan\beta$ is determined with an accuracy of 
$5\%$. Excellent precision of 2-3\% is obtained for the mass of the
otherwise unobservable sneutrino. 
Including NLO effects even allows us
to constrain the parameters of the stop sector. Although the
precision shown in tab.~\ref{tab:ressc1} is rather limited, this could lead to
an important hint concerning the masses of the stops, which, if not already seen, might allow for a 
well-targeted search at the LHC. This could be
another example of LC-LHC interplay~\cite{Weiglein:2004hn}.

Finally, in tab.~\ref{tab:ressc1} we compare the fit results using masses
of the charginos and neutralinos obtained from threshold scans and from the
continuum. For the latter, the accuracy at which the parameters can be determined 
is seen to deteriorate, with errors on the fundamental parameters almost
doubling, 
clearly indicating the need to measure chargino and neutralino masses via
threshold scans.

\begin{table}[tb!]
\renewcommand{\arraystretch}{1.5}
\begin{center}
\begin{tabular}{lr@{}l@{ }l r@{}l@{ }l}\toprule
Parameter & \multicolumn{3}{c}{Threshold fit} & \multicolumn{3}{c}{Continuum fit} \\ 
\midrule
$M_1$ &  $125 $ & $\pm 0.3 $ & $ (\pm 0.7) $ & $125 $ & $\pm 0.6 $ & $ (\pm 1.2)$ \\ 
$M_2$ & $250 $ & $\pm 0.6 $ & $ (\pm 1.3) $  & $250 $ & $\pm 1.6 $ & $ (\pm 3)$ \\ 
$\mu$ &  $180 $ & $\pm 0.4 $ & $ (\pm 0.8) $ & $180 $ & $\pm 0.7 $ & $ (\pm 1.3)$ \\ 
$\tan\beta$ & $10 $ & $ \pm 0.5 $ & $ (\pm 1) $  & $10 $ & $\pm 1.3 $ & $ (\pm 2.6) $ \\
$m_{\tilde{\nu}}$ & $1500 $ & $\pm 24 $ & $ (^{+60}_{-40}) $  & $1500 $ & $\pm 20 $ & $ (\pm 40) $
\\
$\cos\theta_{t}$ & $0 $ & $\pm 0.15 $ & $ (^{+0.4}_{-0.3}) $ & \multicolumn{3}{l}{$\qquad-$} \\ 
$m_{\tilde{t}_1}$ & $400 $ & $^{+180}_{-120} $ & $ (^{\textrm{at limit}}_{\textrm{at
limit}}) $  & \multicolumn{3}{l}{$\qquad-$} \\
$m_{\tilde{t}_2}$ & $800$ & $^{+300}_{-170} $ & $ (^{+1000}_{-290}) $  &
$800$ & $^{+350}_{-220} $ & $  (^{\textrm{at limit}}_{\textrm{at
limit}}$) \\ \bottomrule
\end{tabular}
\end{center}
\caption{Fit results (masses in GeV) for S1, for masses obtained from threshold scans
(threshold fit) 
and from the continuum (continuum fit). Numbers in brackets denote $2\sigma$
errors.\label{tab:ressc1}}
\end{table}

\subsection{Results for scenario 2 \label{sec:scenario2}}

In this scenario, where the $M_2$ parameter is set to 2~TeV, only the light
chargino and three lightest neutralinos will be accessible at the LC.
As input for the fit we therefore use:
\begin{itemize}
\item the masses of the lighter chargino ($\cha^\pm_1$) and neutralinos ($\neu{1},
\neu{2}, \neu{3}$)
\item the light chargino production cross section $\sigma(\cha^+_1\cha^-_1)$
with polarised beams at $\sqrt{s} = 400$ and $500 \gev$
\item the forward-backward asymmetry $A_{FB}$ at $\sqrt{s} = 400$ and $500 \gev$
\item the branching ratio $\mathcal{B}(b\to s\gamma)$.
\end{itemize}
As we again find that the muon anomalous magnetic moment has a negligible impact,
it is not used in the fit.
The input variables, together with errors, namely the assumed experimental precision 
of the prospective LC measurements as well as the theoretical uncertainties, are listed in tab.~\ref{tab:inputsc2}.
While $A_{FB}$ is negligible at LO, the NLO corrections to it are again found to be large. 
\begin{table}[tb!]
\renewcommand{\arraystretch}{1.4}
\begin{center}
\begin{tabular}{@{}lr@{.}lr@{.}lr@{.}lr@{.}l@{}}
\toprule
Observable & \multicolumn{2}{c}{Tree value} & \multicolumn{2}{c}{Loop corr.} & \multicolumn{2}{c}{Error exp.} & \multicolumn{2}{c}{Error th.}\\ 
\midrule
$\mcha{1}$ & \quad$179$ & $1$ & \multicolumn{2}{c}{$-$} & $\quad\quad0$ & $1$ & \multicolumn{2}{c}{$-$} \\
$\mneu{1}$ & $111$ & $1$ & \multicolumn{2}{c}{$-$} & 0 & 2 & \multicolumn{2}{c}{$-$} \\
$\mneu{2}$ & $183$ & $6$ & $0$ & $07$ & 0 & 5 & \quad\,\,0 & 5 \\
$\mneu{3}$ & $194$ & $2$ & $1$ & $9$ & 0 & 5 & 0 & 5 \\
$\sigma(\cha^+_1\cha^-_1)^{400}_{(-0.8,0.6)}$ & $1214$ & $9$ & $-344$ & $7$ & 6 & 0 & 0 & 1
 \\ 
$\sigma(\cha^+_1\cha^-_1)^{400}_{(0.8,-0.6)}$ & $250$ & $6 $ & $-32$ & $4 $ & 2 & 7 & 0 & 1
  \\ 
$\sigma(\cha^+_1\cha^-_1)^{500}_{(-0.8,0.6)}$ & $1079$ & $2 $ & $-194$ & $8 $ & 6 & 0  & 0 & 1
 \\
$\sigma(\cha^+_1\cha^-_1)^{500}_{(0.8,-0.6)}$ & $229$ & $6  $ & $ -8$ & $7 $ & 2 & 7 & 0 & 1
  \\
$A_{FB}^{400}(\%)$ &0&0& 3&0&1&0&0&1\\
$A_{FB}^{500}(\%)$ &0&0& 5&0&1&0&0&1\\
\bottomrule
\end{tabular}
\end{center}
\caption{Observables (masses in GeV, cross sections in fb) used as an input for the fit in S2, as in tab.~\ref{tab:inputsc2}.  
The central value of the theoretical prediction, 
$\mathcal{B}(b\to s\gamma)= 3.3\cdot 10^{-4}$ GeV is also included in the fit. 
See text for details of the calculation and error estimation.\label{tab:inputsc2}}
\end{table}

\begin{table}[tb!]
\renewcommand{\arraystretch}{1.4}
\begin{center}
\begin{tabular}{lr@{}l@{ }l}
\toprule
Parameter & \multicolumn{3}{c}{Fit result} \\ 
\midrule
$M_1$ &  $125$ & $_{-0.6}^{+0.9} $ & $ (_{-1.2}^{+2.1}) $  \\ 
$M_2$ & $2000 $ & $ \pm 200 $ & $ (_{-400}^{+600}) $   \\
$\mu$ &  $180 $ & $\pm 0.2 $ & $ (_{-0.3}^{+0.5}) $ \\
$\tan\beta$ & $10 $ & $\pm 2 $ & $ (_{-4}^{+5}) $  \\
$m_{\tilde{\nu}}$ & \multicolumn{3}{c}{unconstrained}  \\
$\cos\theta_{t}$ & $0$ & $_{-0.09}^{+0.13} $ & $ (^{+0.4}_{-0.3}) $ \\
$m_{\tilde{t}_1}$ & $400$ & $^{+250}_{-50} $ & $ (^{+500}_{-80}) $\\
$m_{\tilde{t}_2}$ & $800$ & $^{+300}_{-200} $ & $ (^{+900}_{-400}) $\\ 
\bottomrule
\end{tabular}
\end{center}
\caption{Fit results (in GeV with the exception of $\tan\beta$ and $\cos\theta_{t}$) for S2, as in tab.~\ref{tab:ressc1}, where numbers
in brackets denote $2\sigma$ errors.\label{tab:ressc2}}
\end{table}

We again fit 8 MSSM parameters: $M_1$, $M_2$, $\mu$, $\tan\beta$,
$m_{\tilde{\nu}}$, $\cos\theta_{t}$, $m_{\tilde{t}_1}$, and
$m_{\tilde{t}_2}$. The impact of other parameters, except the heavy Higgs boson
mass, can be neglected. The results from the fit are given in
tab.~\ref{tab:ressc2}. 
The higgsino and bino mass parameters are well constrained in this scenario
since
bino-like neutralino and all higgsinos are directly accessible. Even though the
winos are not directly accessible, the wino mass parameter $M_2$ can be
constrained
with $10\%$ accuracy at 1$\sigma$ level. An accuracy of $20\%$ is achieved for
$\tan\beta$,
significantly worse than in S1. This can be understood by the fact that the
mixing in S2
between chargino states is weak due to $M_2$ being heavy, and the constraint on
$\tan\beta$
is dependent on this mixing. 
No limits can be derived on the sneutrino mass, due to the Yukawa suppressed
coupling of the higgsino-like $\cha_1^{\pm}$ to the electron
and sneutrino. 
We are however, as shown in tab.~\ref{tab:ressc2}, still able to derive limits
on the stop masses and mixing angle.  

\subsection{Results for scenario 3}
This final scenario features the richest phenomenology of the studied
benchmark scenarios.
As input for the fit we therefore use:
\begin{itemize}
\item the masses of the charginos ($\cha_1^{\pm}$, $\cha_2^{\pm}$) and neutralinos
($\neu{1}, \neu{2}, \neu{3}$)
\item the light chargino production cross section $\sigma(\cha^+_1\cha^-_1)$
with polarised beams at $\sqrt{s} = 400$ and $500 \gev$
\item the forward-backward asymmetry $A_{FB}$ at $\sqrt{s} = 400$ and $500 \gev$
\item the Higgs boson mass, $m_h$
\item the branching ratio $\mathcal{B}(b\to s\gamma)$
\item the anomalous muon magnetic moment
\end{itemize}
Compared to the previous scenarios, these observables are supplemented by the
Higgs
boson mass, $m_h$, calculated using \texttt{FeynHiggs
2.9.1}~\cite{Heinemeyer:1998yj,Heinemeyer:1998np,Degrassi:2002fi,Frank:2006yh}. 
The estimated experimental precision at the LC for $m_h$, taken from
ref.~\cite{AguilarSaavedra:2001rg}, is adopted. We further assume the future theoretical uncertainty on the Higgs boson mass to be 1~GeV~
\cite{Frank:2006yh}. 
As before, the remaining two observables, the branching ratio $\mathcal{B}(b\to s\gamma)$ 
and the anomalous muon magnetic moment are calculated using~{\texttt{micrOmegas 2.4.1}}~\cite{Belanger:2006is,
Belanger:2010gh}, and a projected  
experimental error on the anomalous 
muon magnetic moment of $3.4\cdot10^{-10}$ is employed~\cite{Carey:2009zzb}, which we assume would dominate
over the theoretical uncertainty.
The input variables, together with errors, namely the assumed experimental precision 
of the prospective LC measurements and the theoretical uncertainties, are summarised in tab.~\ref{tab:inputsc3}.
As the sneutrino is now directly accessible, we assume that its mass has been measured and it is therefore not included in the fit.
On the other hand, due to the stronger dependence of the NLO cross-section and forward-backward asymmetry 
on $m_{A^0}$, this is now used as an additional fit prameter. We neglect the remaining theoretical uncertainty on the cross-sections, as it 
is found to be negligible in comparison to the experimental error.

\begin{table}[tb!]
\renewcommand{\arraystretch}{1.5}
\begin{center}
\begin{tabular}{@{}lr@{.}lr@{.}lr@{.}lc@{}}
\toprule
Observable & \multicolumn{2}{c}{Tree value} & \multicolumn{2}{c}{Loop corr.} & \multicolumn{2}{c}{Error exp.} & Error th.\\ 
\midrule
$\mcha{1}$ & \quad$139$ & $3$ & \multicolumn{2}{c}{$-$} & \quad\,\,0 & 1 & $-$\\  
$\mcha{2}$ & $266$ & $2$ & \multicolumn{2}{c}{$-$} & 0 & 5 & $-$\\  
$\mneu{1}$ & $92$ & $8$ & \multicolumn{2}{c}{$-$} & 0 & 2 & $-$\\  
$\mneu{2}$ & $148$ & $5$ & $\qquad2$ & $4$ & 0 & 5 & 0.5\\  
$\mneu{3}$ & $189$ & $7$ & $-7$ & $3$ & 0 & 5 & 0.5\\  
$\sigma(\cha^+_1\cha^-_1)^{400}_{(-0.8,0.6)}$ & $709$ & $7$ & $-85$ & $1$ & 4 & 5 & $-$ \\
 
$\sigma(\cha^+_1\cha^-_1)^{400}_{(0.8,-0.6)}$ & $129$ & $8 $ & $20$ & $0 $ & 2 & 0 & $-$ \\
 
$\sigma(\cha^+_1\cha^-_1)^{500}_{(-0.8,0.6)}$ & $560$ & $0 $ & $ -70$ & $1$ & 4 & 5 & $-$ \\
 
$\sigma(\cha^+_1\cha^-_1)^{500}_{(0.8,-0.6)}$ & $97$ & $1  $ & $ 16$ & $4 $ & 2 & 0 & $-$ 
\\ 
$A_{FB}^{400}(\%)$ & 24&7&$-2$&8& 1&4 & $0.1$\\
$A_{FB}^{500}(\%)$ &39&2&$-5$&8 & 1&5 & $0.1$\\
\bottomrule
\end{tabular}
\end{center}
\caption{Observables (masses in GeV, cross sections in fb) used as an input for the fit in S3, as in tab.~\ref{tab:inputsc1}. 
The central values of the theoretical predictions $ \mathcal{B}(b\to s\gamma)=2.7\cdot 10^{-4}$, 
$\Delta (g_\mu -2)/2=2.4\cdot 10^{-9}$ and $m_h=125$ GeV are also included in the fit. 
See text for details of the calculation and error estimation.\label{tab:inputsc3}}
\end{table}

This means that in scenario 3, we fit to $M_1$, $M_2$,
$\mu$, $\tan\beta$, $\cos\theta_{t}$, $m_{\tilde{t}_1}$, $m_{\tilde{t}_2}$ and $m_{A^0}$.
The results of the fit are collected in
tab.~\ref{tab:ressc3}. The parameters of the electroweak gaugino-higgsino sector
are determined with high precision.
Due to a significant mixing in the stop sector, and the improvement in the fit
quality 
due to the inclusion of the higgs mass, we find that the fit is now also more
sensitive to the mass of the heavy stop. The accuracy is better than $20\%$ for this
particle even though it is far beyond the reach of the LC and also most likely of the LHC. In
addition, in this scenario an upper limit on the mass of the heavy Higgs boson can be placed at 1000~GeV,
at the 2$\sigma$ level. It is the particular sensitivity of the NLO corrections 
to $m_{A^0}$ which presents this unique opportunity to set such an upper bound.

\begin{table}[tb!]
\renewcommand{\arraystretch}{1.5}
\begin{center}
\begin{tabular}{lr@{}l@{ }l}
\toprule
Parameter & \multicolumn{3}{c}{Fit result}  \\ 
\midrule
$M_1$ &  $106$ &$\pm 0.3$ & $(\pm 0.5) $  \\  
$M_2$ & $212$ &$\pm 0.5 $ & $ (\pm 1.0) $   \\  
$\mu$ &  $180$ &$\pm 0.4 $ & $ (\pm 0.9) $  \\  
$\tan\beta$ & $12$ &$\pm 0.3 $ & $ (\pm 0.7) $ \\  
$\cos\theta_{t}$ & $0.15$&$^{+0.08}_{-0.06} $ & $ (^{+0.16}_{-0.09}) $  \\
 
$m_{\tilde{t}_1}$ & $430$&$^{+200}_{-130} $ & $ (^{+300}_{-400}) $  \\  
$m_{\tilde{t}_2}$ & $1520$&$^{+200}_{-300} $ & $ (^{+300}_{-400}) $  \\  
$m_{A^0}$ & \multicolumn{2}{c}{$<650$}  & $ (<1000) $ \\ 
\bottomrule
\end{tabular}
\end{center}
\caption{Fit results (in GeV with the exception of $\tan\beta$ and $\cos\theta_{t}$) for S3, including results for the masses of the heavier stop mass
($m_{\tilde{t}_2}$) and 
the pseudoscalar higgs boson ($m_{A^0}$).}\label{tab:ressc3}
\end{table}


\section{Conclusions}\label{sec:conc}
The evidence for the Higgs boson and dark matter,
when examined in the context of supersymmetry, suggests the possibility of a light
$\mu$ and $M_1$.
We have extended previous analyses, which fitted observables
for chargino
production at the LC to extract fundamental MSSM parameters, by incorporating
NLO corrections. The loop corrections are calculated for all observables fitted,
namely
the polarised cross-sections and forward backward asymmetry for chargino
production as well as the $\cha_1^{\pm},\cha_2^{\pm}$
and $\neu{1}, \neu{2}, \neu{3}$ masses, in an on-shell scheme which facilitates
the extension to the complex case. We have fitted these observables in three
complementary scenarios.

On including NLO corrections, we found that when  $M_1$, $M_2$ and $\mu$ are
light they can be determined to percent-level accuracy, and $\tan\beta$ to $<5\%$.
Further we showed that if masses of the charginos and neutralinos are obtained from the continuum
as opposed to via threshold scans, the uncertainty on the fundamental parameters would almost double, reinforcing
the importance of threshold scans for mass measurements.
As a heavy $M_2$ is still a viable possibility, we also considered
$M_2=2000\gev$, and found that the sensitivity to $M_2$ is approximately $10\%$.
However in this case the error on $\tan\beta$, dependent on the degree of mixing in the chargino
sector, increases to $\sim20\%$.
Note that the inclusion of $\mathcal{B}(b\to s\gamma)$ in the fit, in combination with the use of
masses determined via threshold scanning, was seen to 
improve the sensitivity to the stop sector.

The final scenario we considered was compatible with the latest Higgs results.
Here we found that including $\mathcal{B}(b\to s\gamma)$, $\Delta (g_\mu
-2)/2$ and $m_h$ in the fit, 
as well as the fact that there was significant mixing in the stop sector, helped to obtain
an accuracy better than $20\%$ on the mass of the heavy stop,
even though this particle is far beyond the reach of the LC and also most likely of the LHC. We 
also included $m_{A^0}$ in the fit, and found that, due to the particular sensitivity of 
the NLO corrections to $m_{A^0}$, it would even be possible
to place a 2$\sigma$ upper bound on this parameter of 1000~GeV.
In summary, we have shown that incorporating NLO corrections is required for the
precise 
determination of the fundamental parameters of the chargino and neutralino 
sector at the LC, and could further provide sensitivity to the parameters
describing particles
which contribute via loop corrections.

\section*{Acknowledgements}
The authors gratefully acknowledge support of the DFG through the grant SFB 676,
``Particles, Strings, and the Early Universe'', as well as the Helmholtz
Alliance, ``Physics at the
 Terascale''.
This work was also partially supported by the Polish National Science Centre 
under research grant DEC-2011/01/M/ST2/02466 and the MICINN, Spain, under contract 
FPA2010-17747; Consolider-Ingenio  CPAN CSD2007- 00042. KR thanks as well the 
Comunidad de Madrid through Proyecto HEPHACOS S2009/ESP-1473 and the European 
Commission under contract PITN-GA-2009-237920.

\bibliography{BKMRW}{}

\providecommand{\href}[2]{#2}\begingroup\raggedright\begin{thebibliography}{10}

\bibitem{AguilarSaavedra:2001rg}
{\bf ECFA/DESY LC Physics Working Group} Collaboration, J.~Aguilar-Saavedra
  et~al. \href{http://xxx.lanl.gov/abs/hep-ph/0106315}{{\tt hep-ph/0106315}}.

\bibitem{Abe:2001gc}
{\bf ACFA Linear Collider Working Group} Collaboration, K.~Abe et~al.
  \href{http://xxx.lanl.gov/abs/hep-ph/0109166}{{\tt hep-ph/0109166}}.

\bibitem{Abe:2001wn}
{\bf American Linear Collider Working Group} Collaboration, T.~Abe et~al.
  \href{http://xxx.lanl.gov/abs/hep-ex/0106056}{{\tt hep-ex/0106056}}.

\bibitem{BrauJames:2007aa}
{\bf ILC} Collaboration, E.~Brau, James et~al.
  \href{http://xxx.lanl.gov/abs/0712.1950}{{\tt arXiv:0712.1950}}.

\bibitem{Djouadi:2007ik}
{\bf ILC} Collaboration, G.~Aarons et~al.
  \href{http://xxx.lanl.gov/abs/0709.1893}{{\tt arXiv:0709.1893}}.

\bibitem{Goldberg:1983nd}
H.~Goldberg {\em Phys.Rev.Lett.} {\bf 50} (1983) 1419.

\bibitem{Ellis:1983ew}
J.~R. Ellis, J.~Hagelin, D.~V. Nanopoulos, K.~A. Olive, and M.~Srednicki {\em
  Nucl.Phys.} {\bf B238} (1984) 453--476.

\bibitem{Hall:2011aa}
L.~J. Hall, D.~Pinner, and J.~T. Ruderman {\em JHEP} {\bf 1204} (2012) 131,
  [\href{http://xxx.lanl.gov/abs/1112.2703}{{\tt arXiv:1112.2703}}].

\bibitem{Brummer:2011yd}
F.~Brummer and W.~Buchmuller {\em JHEP} {\bf 1107} (2011) 010,
  [\href{http://xxx.lanl.gov/abs/1105.0802}{{\tt arXiv:1105.0802}}].

\bibitem{Chatrchyan:2012pka}
{\bf CMS Collaboration} Collaboration, S.~Chatrchyan et~al. {\em JHEP} {\bf
  1211} (2012) 147, [\href{http://xxx.lanl.gov/abs/1209.6620}{{\tt
  arXiv:1209.6620}}].

\bibitem{Aad:2012hba}
{\bf ATLAS Collaboration} Collaboration, G.~Aad et~al. {\em Phys.Lett.} {\bf
  B718} (2013) 841--859, [\href{http://xxx.lanl.gov/abs/1208.3144}{{\tt
  arXiv:1208.3144}}].

\bibitem{Fittino}
P.~Bechtle, T.~Bringmann, K.~Desch, H.~Dreiner, M.~Hamer, et~al. {\em JHEP}
  {\bf 1206} (2012) 098, [\href{http://xxx.lanl.gov/abs/1204.4199}{{\tt
  arXiv:1204.4199}}].

\bibitem{ILCDBD}
{\bf ILC} Collaboration, H.~Baer et~al., ``{International Linear Collider
  Technical Design Report - Volume 1: Physics at the International Linear
  Collider}.'' 2012.

\bibitem{Oller:2005xg}
W.~Oller, H.~Eberl, and W.~Majerotto {\em Phys.Rev.} {\bf D71} (2005) 115002,
  [\href{http://xxx.lanl.gov/abs/hep-ph/0504109}{{\tt hep-ph/0504109}}].

\bibitem{Fritzsche:2004nf}
T.~Fritzsche and W.~Hollik {\em Nucl.Phys.Proc.Suppl.} {\bf 135} (2004)
  102--106, [\href{http://xxx.lanl.gov/abs/hep-ph/0407095}{{\tt
  hep-ph/0407095}}].

\bibitem{Kilian:2006cj}
W.~Kilian, J.~Reuter, and T.~Robens {\em Eur.Phys.J.} {\bf C48} (2006)
  389--400, [\href{http://xxx.lanl.gov/abs/hep-ph/0607127}{{\tt
  hep-ph/0607127}}].

\bibitem{Robens:2008sa}
T.~Robens, J.~Kalinowski, K.~Rolbiecki, W.~Kilian, and J.~Reuter {\em Acta
  Phys.Polon.} {\bf B39} (2008) 1705--1714,
  [\href{http://xxx.lanl.gov/abs/0803.4161}{{\tt arXiv:0803.4161}}].

\bibitem{bfmw}
A.~Bharucha, A.~Fowler, G.~Moortgat-Pick, and G.~Weiglein
  \href{http://xxx.lanl.gov/abs/1211.3134}{{\tt arXiv:1211.3134}}.

\bibitem{Fritzsche:2005}
T.~Fritzsche.
\newblock PhD thesis, Universitaet Karlsruhe, 2005.

\bibitem{Fowler:2009ay}
A.~Fowler and G.~Weiglein {\em JHEP} {\bf 1001} (2010) 108,
  [\href{http://xxx.lanl.gov/abs/0909.5165}{{\tt arXiv:0909.5165}}].

\bibitem{AlisonsThesis}
A.~Fowler.
\newblock PhD thesis, Durham University, 2010.

\bibitem{Chatterjee:2011wc}
A.~Chatterjee, M.~Drees, S.~Kulkarni, and Q.~Xu
  \href{http://xxx.lanl.gov/abs/1107.5218}{{\tt arXiv:1107.5218}}.

\bibitem{Heinemeyer:2011gk}
S.~Heinemeyer, F.~von~der Pahlen, and C.~Schappacher {\em Eur.Phys.J.} {\bf
  C72} (2012) 1892, [\href{http://xxx.lanl.gov/abs/1112.0760}{{\tt
  arXiv:1112.0760}}].

\bibitem{Bharucha:2012re}
A.~Bharucha, S.~Heinemeyer, F.~von~der Pahlen, and C.~Schappacher
  \href{http://xxx.lanl.gov/abs/1208.4106}{{\tt arXiv:1208.4106}}.

\bibitem{Haber:1984rc}
H.~E. Haber and G.~L. Kane {\em Phys.Rept.} {\bf 117} (1985) 75--263.

\bibitem{Choi:2000ta}
S.~Choi, A.~Djouadi, M.~Guchait, J.~Kalinowski, H.~Song, et~al. {\em
  Eur.Phys.J.} {\bf C14} (2000) 535--546,
  [\href{http://xxx.lanl.gov/abs/hep-ph/0002033}{{\tt hep-ph/0002033}}].

\bibitem{Kublbeck:1990xc}
J.~Kublbeck, M.~Bohm, and A.~Denner {\em Comput.Phys.Commun.} {\bf 60} (1990)
  165--180.

\bibitem{Denner:1992vza}
A.~Denner, H.~Eck, O.~Hahn, and J.~Kublbeck {\em Nucl.Phys.} {\bf B387} (1992)
  467--484.

\bibitem{FAorig}
J.~Kublbeck, H.~Eck, and R.~Mertig {\em Nucl.Phys.Proc.Suppl.} {\bf 29A} (1992)
  204--208.

\bibitem{Hahn:2000kx}
T.~Hahn {\em Comput.Phys.Commun.} {\bf 140} (2001) 418--431,
  [\href{http://xxx.lanl.gov/abs/hep-ph/0012260}{{\tt hep-ph/0012260}}].

\bibitem{Hahn:2001rv}
T.~Hahn and C.~Schappacher {\em Comput.Phys.Commun.} {\bf 143} (2002) 54--68,
  [\href{http://xxx.lanl.gov/abs/hep-ph/0105349}{{\tt hep-ph/0105349}}].

\bibitem{Hahn:1998yk}
T.~Hahn and M.~Perez-Victoria {\em Comput.Phys.Commun.} {\bf 118} (1999)
  153--165, [\href{http://xxx.lanl.gov/abs/hep-ph/9807565}{{\tt
  hep-ph/9807565}}].

\bibitem{FormCalc2}
T.~Hahn {\em Comput.Phys.Commun.} {\bf 178} (2008) 217--221,
  [\href{http://xxx.lanl.gov/abs/hep-ph/0611273}{{\tt hep-ph/0611273}}].

\bibitem{FormCalc3}
T.~Hahn and M.~Rauch {\em Nucl.Phys.Proc.Suppl.} {\bf 157} (2006) 236--240,
  [\href{http://xxx.lanl.gov/abs/hep-ph/0601248}{{\tt hep-ph/0601248}}].

\bibitem{DRED}
W.~Siegel {\em Phys.Lett.} {\bf B84} (1979) 193.

\bibitem{DRED2}
W.~Siegel {\em Phys.Lett.} {\bf B94} (1980) 37.

\bibitem{0503129}
D.~Stockinger {\em JHEP} {\bf 0503} (2005) 076,
  [\href{http://xxx.lanl.gov/abs/hep-ph/0503129}{{\tt hep-ph/0503129}}].

\bibitem{delAguila:1998nd}
F.~{del Aguila}, A.~Culatti, R.~{Munoz Tapia}, and M.~Perez-Victoria {\em
  Nucl.Phys.} {\bf B537} (1999) 561--585,
  [\href{http://xxx.lanl.gov/abs/hep-ph/9806451}{{\tt hep-ph/9806451}}].

\bibitem{Lahanas:1993ib}
A.~Lahanas, K.~Tamvakis, and N.~Tracas {\em Phys.Lett.} {\bf B324} (1994)
  387--396, [\href{http://xxx.lanl.gov/abs/hep-ph/9312251}{{\tt
  hep-ph/9312251}}].

\bibitem{Pierce:1993gj}
D.~Pierce and A.~Papadopoulos {\em Phys.Rev.} {\bf D50} (1994) 565--570,
  [\href{http://xxx.lanl.gov/abs/hep-ph/9312248}{{\tt hep-ph/9312248}}].

\bibitem{Pierce:1994ew}
D.~Pierce and A.~Papadopoulos {\em Nucl.Phys.} {\bf B430} (1994) 278--294,
  [\href{http://xxx.lanl.gov/abs/hep-ph/9403240}{{\tt hep-ph/9403240}}].

\bibitem{Eberl:2001eu}
H.~Eberl, M.~Kincel, W.~Majerotto, and Y.~Yamada {\em Phys.Rev.} {\bf D64}
  (2001) 115013, [\href{http://xxx.lanl.gov/abs/hep-ph/0104109}{{\tt
  hep-ph/0104109}}].

\bibitem{Fritzsche:2002bi}
T.~Fritzsche and W.~Hollik {\em Eur.Phys.J.} {\bf C24} (2002) 619--629,
  [\href{http://xxx.lanl.gov/abs/hep-ph/0203159}{{\tt hep-ph/0203159}}].

\bibitem{Oller:2003ge}
W.~Oller, H.~Eberl, W.~Majerotto, and C.~Weber {\em Eur.Phys.J.} {\bf C29}
  (2003) 563--572, [\href{http://xxx.lanl.gov/abs/hep-ph/0304006}{{\tt
  hep-ph/0304006}}].

\bibitem{Drees:2006um}
M.~Drees, W.~Hollik, and Q.~Xu {\em JHEP} {\bf 0702} (2007) 032,
  [\href{http://xxx.lanl.gov/abs/hep-ph/0610267}{{\tt hep-ph/0610267}}].

\bibitem{Schofbeck:2006gs}
R.~Schofbeck and H.~Eberl {\em Phys.Lett.} {\bf B649} (2007) 67--72,
  [\href{http://xxx.lanl.gov/abs/hep-ph/0612276}{{\tt hep-ph/0612276}}].

\bibitem{Schofbeck:2007ib}
R.~Schofbeck and H.~Eberl {\em Eur.Phys.J.} {\bf C53} (2008) 621--626,
  [\href{http://xxx.lanl.gov/abs/0706.0781}{{\tt arXiv:0706.0781}}].

\bibitem{Rolbiecki:2007se}
K.~Rolbiecki and J.~Kalinowski {\em Phys.Rev.} {\bf D76} (2007) 115006,
  [\href{http://xxx.lanl.gov/abs/0709.2994}{{\tt arXiv:0709.2994}}].

\bibitem{Eberl:2005ay}
H.~Eberl, T.~Gajdosik, W.~Majerotto, and B.~Schrausser {\em Phys.Lett.} {\bf
  B618} (2005) 171--181, [\href{http://xxx.lanl.gov/abs/hep-ph/0502112}{{\tt
  hep-ph/0502112}}].

\bibitem{Osland:2007xw}
P.~Osland and A.~Vereshagin {\em Phys.Rev.} {\bf D76} (2007) 036001,
  [\href{http://xxx.lanl.gov/abs/0704.2165}{{\tt arXiv:0704.2165}}].

\bibitem{Desch:2006xp}
K.~Desch, J.~Kalinowski, G.~Moortgat-Pick, K.~Rolbiecki, and W.~Stirling {\em
  JHEP} {\bf 0612} (2006) 007,
  [\href{http://xxx.lanl.gov/abs/hep-ph/0607104}{{\tt hep-ph/0607104}}].

\bibitem{James:1975dr}
F.~James and M.~Roos {\em Comput.Phys.Commun.} {\bf 10} (1975) 343--367.

\bibitem{minuit}
F.~James {\em CERN Program Library Long Writeup D506} (1994).

\bibitem{atlas:2012rz}
{\bf ATLAS Collaboration} Collaboration, G.~Aad et~al.
  \href{http://xxx.lanl.gov/abs/1208.0949}{{\tt arXiv:1208.0949}}.

\bibitem{cms:2012jx}
{\bf CMS Collaboration} Collaboration, S.~Chatrchyan et~al.
  \href{http://xxx.lanl.gov/abs/1207.1798}{{\tt arXiv:1207.1798}}.

\bibitem{Heinemeyer:1998yj}
S.~Heinemeyer, W.~Hollik, and G.~Weiglein {\em Comput.Phys.Commun.} {\bf 124}
  (2000) 76--89, [\href{http://xxx.lanl.gov/abs/hep-ph/9812320}{{\tt
  hep-ph/9812320}}].

\bibitem{Heinemeyer:1998np}
S.~Heinemeyer, W.~Hollik, and G.~Weiglein {\em Eur.Phys.J.} {\bf C9} (1999)
  343--366, [\href{http://xxx.lanl.gov/abs/hep-ph/9812472}{{\tt
  hep-ph/9812472}}].

\bibitem{Degrassi:2002fi}
G.~Degrassi, S.~Heinemeyer, W.~Hollik, P.~Slavich, and G.~Weiglein {\em
  Eur.Phys.J.} {\bf C28} (2003) 133--143,
  [\href{http://xxx.lanl.gov/abs/hep-ph/0212020}{{\tt hep-ph/0212020}}].

\bibitem{Frank:2006yh}
M.~Frank, T.~Hahn, S.~Heinemeyer, W.~Hollik, H.~Rzehak, et~al. {\em JHEP} {\bf
  0702} (2007) 047, [\href{http://xxx.lanl.gov/abs/hep-ph/0611326}{{\tt
  hep-ph/0611326}}].

\bibitem{ATLAS:2012gk}
{\bf ATLAS Collaboration} Collaboration, G.~Aad et~al. {\em Phys.Lett.} {\bf
  B716} (2012) 1--29, [\href{http://xxx.lanl.gov/abs/1207.7214}{{\tt
  arXiv:1207.7214}}].

\bibitem{CMS:2012gu}
{\bf CMS Collaboration} Collaboration, S.~Chatrchyan et~al. {\em Phys.Lett.}
  {\bf B716} (2012) 30--61, [\href{http://xxx.lanl.gov/abs/1207.7235}{{\tt
  arXiv:1207.7235}}].

\bibitem{Belanger:2006is}
G.~Belanger, F.~Boudjema, A.~Pukhov, and A.~Semenov {\em Comput.Phys.Commun.}
  {\bf 176} (2007) 367--382,
  [\href{http://xxx.lanl.gov/abs/hep-ph/0607059}{{\tt hep-ph/0607059}}].

\bibitem{Belanger:2010gh}
G.~Belanger, F.~Boudjema, P.~Brun, A.~Pukhov, S.~Rosier-Lees, et~al. {\em
  Comput.Phys.Commun.} {\bf 182} (2011) 842--856,
  [\href{http://xxx.lanl.gov/abs/1004.1092}{{\tt arXiv:1004.1092}}].

\bibitem{Komatsu:2010fb}
{\bf WMAP Collaboration} Collaboration, E.~Komatsu et~al. {\em
  Astrophys.J.Suppl.} {\bf 192} (2011) 18,
  [\href{http://xxx.lanl.gov/abs/1001.4538}{{\tt arXiv:1001.4538}}].

\bibitem{O'Leary:2010af}
{\bf SuperB Collaboration} Collaboration, B.~O'Leary et~al.
  \href{http://xxx.lanl.gov/abs/1008.1541}{{\tt arXiv:1008.1541}}.

\bibitem{Weiglein:2004hn}
{\bf LHC/LC Study Group} Collaboration, G.~Weiglein et~al. {\em Phys.Rept.}
  {\bf 426} (2006) 47--358, [\href{http://xxx.lanl.gov/abs/hep-ph/0410364}{{\tt
  hep-ph/0410364}}].

\bibitem{Carey:2009zzb}
R.~Carey, K.~Lynch, J.~Miller, B.~Roberts, W.~Morse, et~al.

\end{thebibliography}\endgroup
\bibliographystyle{JHEP}
\end{document}